\documentclass[preprint,nofootinbib,showpacs,superscriptaddress,aps,prc]{revtex4-1}
\usepackage{graphicx}
\usepackage{amstext}
\usepackage{amssymb}
\usepackage[usenames]{color}

\renewcommand{\Re}{\mathop{\mathrm{Re}}\nolimits}
\renewcommand{\Im}{\mathop{\mathrm{Im}}\nolimits}

\begin{document}
\title{Proton-lambda correlation functions at the LHC \\ with account for residual correlations}

\author{V.~M.~Shapoval} 
\author{Yu.~M.~Sinyukov}
\author{V.~Yu.~Naboka}
\affiliation{Bogolyubov~Institute~for~Theoretical~Physics, Metrolohichna str.~14b, 03680 Kiev, Ukraine }

\begin{abstract}
The theoretical analysis of $\bar{p}-\Lambda \oplus p-\bar{\Lambda}$ 
correlation function in 10\% most central Au+Au collisions at RHIC energy $\sqrt{s_{NN}}=200$~GeV shows that the contribution of
residual correlations is the necessary factor to obtain a satisfactory description of
the experimental data.
A neglecting of the residual correlation effect, leads to unrealistically low source radius, about 2 times
smaller than the corresponding value for $p-\Lambda \oplus \bar{p}-\bar{\Lambda}$
case, when one 
fits the experimental correlation function within Lednick\'y-Lyuboshitz
analytical model.  
Recently an approach 
accounting effectively for residual correlations 
for the baryon-antibaryon correlation function was proposed,
and a good RHIC data description was reached with the source radius
extracted from the hydrokinetic model (HKM). The $\bar{p}-\Lambda$ scattering
length, as well as the parameters characterizing the residual correlation effect
--- annihilation dip amplitude and its inverse width --- were extracted 
from the corresponding fit.
In this paper we use these extracted values and simulated in HKM source functions
for Pb+Pb collisions at the LHC energy $\sqrt{s_{NN}}=2.76$~TeV to predict
the corresponding $p\Lambda$ and $p\bar{\Lambda}$ correlation functions.   
\end{abstract}

\maketitle



\section{Introduction}

The study of $p \Lambda$ correlation functions (CF)
(along with the study of other two-particle correlations such as $\pi\pi$, $KK$)
allows one to obtain the information about the 
character of evolution of the matter formed in relativistic nuclear collisions,
in particular about the character of collective flow. 

Another great possibility which is open for a researcher in baryon-(anti)baryon
correlation analysis is the study of strong interaction between the
particles of different sorts with Final State Interaction (FSI) correlation 
technique \cite{Led,fsi,FsiSin}. 
Since LHC produces copiously hadrons of different species, including
multi-strange, charmed and beauty ones, 
the advantage of this approach is the ability to analyze the interactions
even in exotic particle pairs, hardly achieved by other means.

Fitting the experimental correlation
function with some analytical formula, e.~g.~Lednick\'y-Lyuboshitz model \cite{Led},
allows one to extract the quantities describing both the interaction in corresponding pairs
and the particle emission region size.
Of course, if the source function~$S({\bf r}^*)$, describing the spatial structure of the pair emission, is known, an extraction of unknown interaction parameters becomes more reliable. The corresponding spatial structure can be obtained from realistic collision model that simulates the evolution of the system formed in high energy heavy ion collisions.
In the recent paper~\cite{pLamOur} the hybrid variant of the hydrokinetic model 
(HKM)~\cite{HKM,Boltz,Kaon} was used for this purpose. 
The choice of HKM is highly reliable since it is known to provide a successful simultaneous description of a wide class of bulk observables in nuclear collision 
experiments at RHIC and LHC \cite{Uniform}.
The model also reproduces well~\cite{Sf} the source functions for pion and kaon pairs
in Au+Au collisions at the top RHIC energy~\cite{Phenix}, including 
non-Gaussian tails observed in certain experimental source function projections. 

In~\cite{pLamOur} in order to extract unknown $\bar{p}\Lambda$ scattering length, 
the corresponding experimental correlation function for 10\% most central Au+Au collisions 
at top RHIC energy, measured by STAR Collaboration~\cite{Star}, was fitted with the Lednick\'y-Lyuboshitz formula using the effective source radius $r_0$ extracted from the HKM source function. 
It was also found that $r_0$ values obtained in HKM for baryon-baryon and baryon-antibaryon
correlations are expectedly close, while in the STAR experimental analysis~\cite{Star}, 
where the source radii were considered as free fit parameters, the extracted $\bar{p} \Lambda$
radius was $\sim 2$ times smaller than the $p \Lambda$ one.
One can assume that this apparent difference is due to neglect of residual correlations in~\cite{Star}.
Such correlations can exist between secondary protons and lambdas
if their parents were correlated (or the parent of one particle was correlated with
another particle in the pair).
For taking into account the residual correlation contribution to the baryon-antibaryon CF, 
a modified analytical formula was introduced in~\cite{pLamOur}.
As a result, a good description of the experimental $\bar{p} \Lambda$ correlation function
is obtained and the $\bar{p} \Lambda$ spin-averaged scattering length is extracted from the 
corresponding fit.

In the present paper we are going to apply the method developed in~\cite{pLamOur}
for description of the RHIC data to predict the $p \Lambda$ and $\bar{p} \Lambda$
correlation functions in the 5\% most central Pb+Pb collisions at the LHC
energy $\sqrt{s_{NN}}=2.76$~TeV, utilizing the results of Ref.~\cite{pLamOur} as a starting
point.

\section{Formalism}

The experimental correlation functions presented in~\cite{Star} are purity corrected.
They are obtained from the measured ones as
\begin{equation}\label{pur}
C_{\mathrm{corr}}(k^{*}) = \frac{C_{\mathrm{meas}}(k^{*})-1}{\lambda(k^{*})}+1,
\end{equation}
where $C_{\mathrm{meas}}(k^{*})$ and $C_{\mathrm{corr}}(k^{*})$ are the measured and the corrected
CF respectively, and $\lambda(k^{*})$ is pair purity. The latter is defined
as the fraction of pairs consisting of primary, correctly identified particles.

As well as in~\cite{pLamOur} we model the purity corrected LHC baryon-baryon correlation
function with Lednick\'y-Lyuboshitz analytical formula~\cite{Led}:
\begin{eqnarray}\label{LL}
&& C(k^*) = 1 + \sum_S \rho_S\left[\frac12\left|\frac{f^S(k^*)}{r_0}
\right|^2\left(1-\frac{d_0^S}{2\sqrt\pi r_0}\right)\right. + \nonumber\\
&&\left.\frac{2\Re f^S(k^*)}{\sqrt\pi r_0}F_1(2k^* r_0)-
\frac{\Im f^S(k^*)}{r_0}F_2(2k^* r_0)\right],
\end{eqnarray}
where $F_1(z) = \int_0^z dx e^{x^2 - z^2}/z$ and $F_2(z) = (1-e^{-z^2})/z$.
This formula is derived starting from the basic equation
$ C(k^{*}) = \left\langle \left| \Psi^{S}_{- \textbf{k}^{*}}(\textbf{r}^{*})\right|^{ 2}\right\rangle$, where the wave function $\Psi_{- \textbf{k}^{*}}^{S}$ 
represents the stationary solution of the scattering problem with the opposite sign of the vector $\textbf{k}^{*}$. The angle brackets mean averaging over the total spin $S$ and the distribution of 
the relative distances $S(\textbf{r}^{*})$. Since typically the source radius can be considered much 
larger than the range of the strong interaction potential, $\Psi_{- \textbf{k}^{*}}^{S}$
can be approximated at small $k^*$ by the s-wave solution in the outer region:
\begin{equation}
\Psi_{- \textbf{k}^{*}}^{S} (\textbf{r}^{*}) = e^{-i\textbf{k}^{*} \cdot \textbf{r}^{*}}+ \frac{f^{S}(k^{*})}{r^{*}} e^{ik^{*} \cdot r^{*}}.
\end{equation}
The scattering amplitude $f^{S}(k^{*})$ here is taken in the effective range approximation
\begin{equation}\label{l5}
f^{S}(k^{*}) = \left( \frac{1}{f^{S}_{0}} + \frac{1}{2} d^{S}_{0} k^{*2} - i k^{*} \right)^{-1},
\end{equation}
where $f^{S}_{0}$ is the scattering length and $d^{S}_{0}$ is the effective radius for a given 
total spin $S=1$ or $S=0$.
The singlet and triplet weights $\rho_i$ for unpolarized particles (supposing the polarization $P=0$) are 
in $\rho_0 = 1/4 (1-P^2)=1/4$ and $\rho_1 = 1/4 (3+P^2)=3/4$ correspondingly.

As for the baryon-antibaryon case, following~\cite{pLamOur}, we fit the experimentally measured 
correlation function~$C_{uncorr}(k^{*})$ with the following analytical expression 
\begin{equation}\label{cfunc}
C_{\mathrm{uncorr}}(k^{*}) = 1 + \lambda(k^{*}) (C(k^{*})-1) + \alpha(k^*) (C_{\mathrm{res}}(k^*)-1),
\end{equation}
where $\lambda(k^{*})$ is purity or the fraction of correctly identified pairs
consisting of primary particles, 
$C(k^{*})$ is ``true'' correlation function approximated by Eq.~(\ref{LL}),
$\alpha(k^*)$ is the fraction of secondary 
particles which are residually correlated, $\alpha(k^*)=\tilde{\alpha}(1-\lambda(k^*))$
and $C_{\mathrm{res}}(k^*)$ is the residual correlation contribution.
The latter is taken in the Gaussian form~\cite{pLamOur,StarPRL}
\begin{equation}\label{G}
C_{\mathrm{res}}(k^*)=1-\tilde{\beta} e^{-4k^{*2}R^2},
\end{equation} 
where $\tilde{\beta} = A > 0$ is the annihilation (wide) dip amplitude and $R \ll r_0$ is the dip inverse width.  
Since $\tilde{\alpha}$ and $\tilde{\beta}$ enter (\ref{cfunc}) only as a product
$\tilde{\alpha}\tilde{\beta}$, the latter is treated as a single parameter $\beta$
at fitting.

The source radii $r_0$ in both $p\Lambda$ and $\bar{p}\Lambda$ cases are extracted from the Gaussian fit $S_{\mathrm{fit}}(r^*) = (2\sqrt{\pi}r_0)^{-3} e^{-\frac{r^{*2}}{4 r_0^2}}$
to the angle averaged source function, 
\begin{equation}
S(r^*)=1/(4\pi)\int_0^{2\pi} \int_0^{\pi} S(r^*,\theta,\phi) \sin\theta d\theta d\phi,
\label{sf}
\end{equation}
calculated in the hydrokinetic model~\cite{HKM,Boltz,Kaon}.
The latter simulates the evolution of the matter in relativistic nuclear collision
as consisting of two stages --- the hydrodynamic expansion of the matter being
in local thermal and chemical equilibrium and gradual system decoupling,
beginning when the equilibrium is lost.
The first stage is described within ideal hydrodynamics and for the second one
the hydrokinetic approach is utilized, based on the Boltzmann equations in the integral form,
with switching to UrQMD cascade at a space-like hypersurface.
In current study HKM is taken
in its simplified hybrid form~\cite{Uniform} with sudden switch 
from hydrodynamic evolution to the cascade at the hadronization
hypersurface defined by the isotherm $T=165$~MeV.

The model output consists of generated particle momenta and coordinates,
that are further used to build different observables.
The considered angle-averaged source function histograms are filled using the following procedure
(here $r^{*}$ is the particle spatial separation in the pair rest frame) 
\begin{equation}\label{fill}
S({r}^{*(k)})= \frac{\sum_{n=1}^{N_{\mathrm{ev}}}\sum_{i_1^n,i_2^n}
    [\delta_{\Delta}
    (r^{*(k)}-r^*_{i_1^n}+r^*_{i_2^n})/(4\pi(r^*_{i_1^n}-r^*_{i_2^n})^2\Delta)]}{\sum_{n=1}^{N_{\mathrm{ev}}}\sum_{i_1^n,i_2^n}1}
\end{equation}
Here ${r}_{i_1^n}^*$ and ${r}_{i_2^n}^*$ are the pair rest frame $r$-coordinates 
of particles 1 and 2 produced in the $n-$th event, ${r}^{*(k)}$ is the $r$-coordinate 
of the $k$-th histogram bin center, the function $\delta_\Delta(x)=1$ 
if $|x|<\Delta/2$ and $0$ otherwise, and $\Delta$ is the size of the 
histogram bin.

\section{Results and discussion}

The initial conditions (IC) for HKM calculations simulating the considered case of
5\% most central Pb+Pb collisions at the LHC energy $\sqrt{s_{NN}}=2.76$~TeV
are described in detail in~\cite{Uniform,Shap}.
We assume longitudinal boost invariance, so that the IC are specified in the plane transverse to the beam axis only.
The transverse energy density profile $\epsilon_i(\textbf{r}_T)$ at the starting time 
$\tau_i=0.1$~fm/$c$ corresponds to Monte Carlo Glauber model and is calculated in \textsc{GLISSANDO} 
code~\cite{Gliss}, where the overall scale factor $\epsilon_0$, being the maximal initial energy 
density, is fixed basing on the experimental mean charged particle multiplicity.
The initial transverse flow in present calculations is absent.

\begin{figure}[t]
\centering
\includegraphics[bb=0 0 567 409,width=0.88\textwidth]{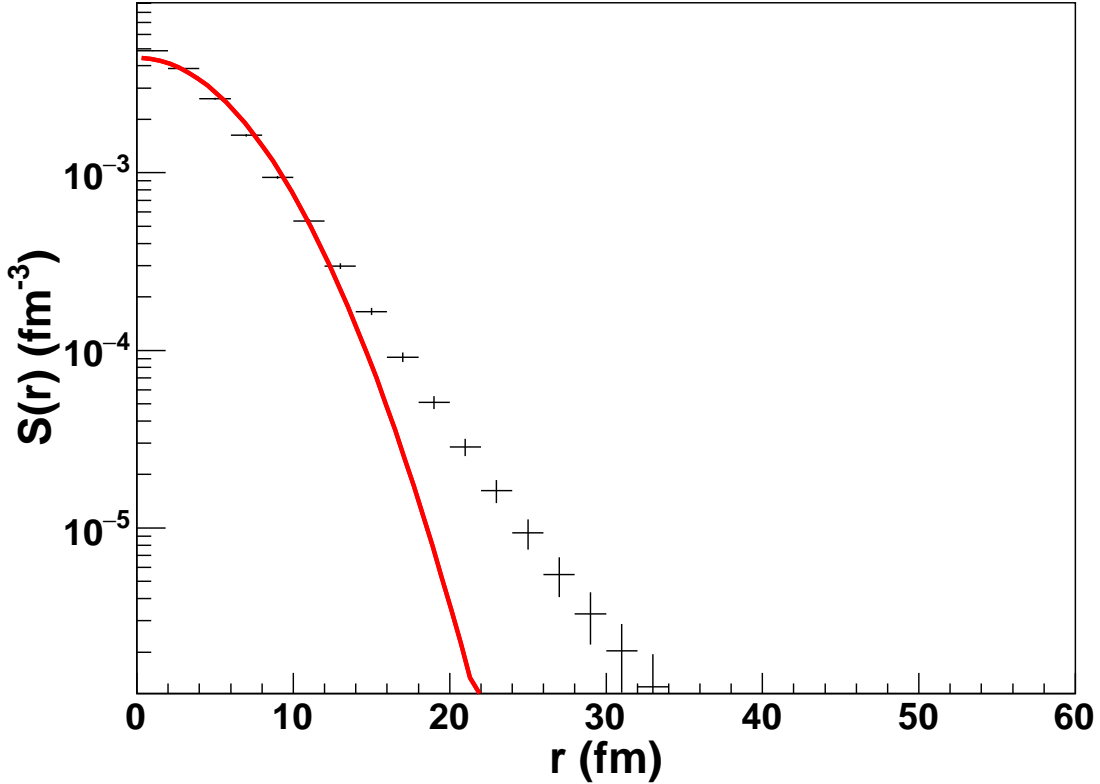}
  \caption{The $p\Lambda$ angle averaged source function calculated in HKM (markers) and the Gaussian fit to it (line).
The calculations correspond to 5$\%$ most central Pb+Pb collisions at LHC energy $\sqrt{s_{NN}}=2.76$~TeV, 
in pseudorapidity range $|\eta|<0.8$. Proton $p_T$ range is $0.7<p_T<4$~GeV/$c$,
and for lambdas $0.7<p_T<5$~GeV/$c$.
The extracted $r_0=3.76$~fm.
\label{sfav}}
\end{figure}

In Fig.~\ref{sfav} one can see the angle averaged $p\Lambda$ source function $S(r^*)$
obtained in hydrokinetic model for considered LHC collisions together
with the Gaussian fit to it. The source radius value extracted from this fit is $r_0=3.76$~fm,
which is about 1.15 times larger than for the RHIC case. 
The $\bar{p}\Lambda$ source function fitting results in the same source radius value.

\begin{figure}[t]
\centering
\includegraphics[bb=0 0 567 409, width=0.9\textwidth]{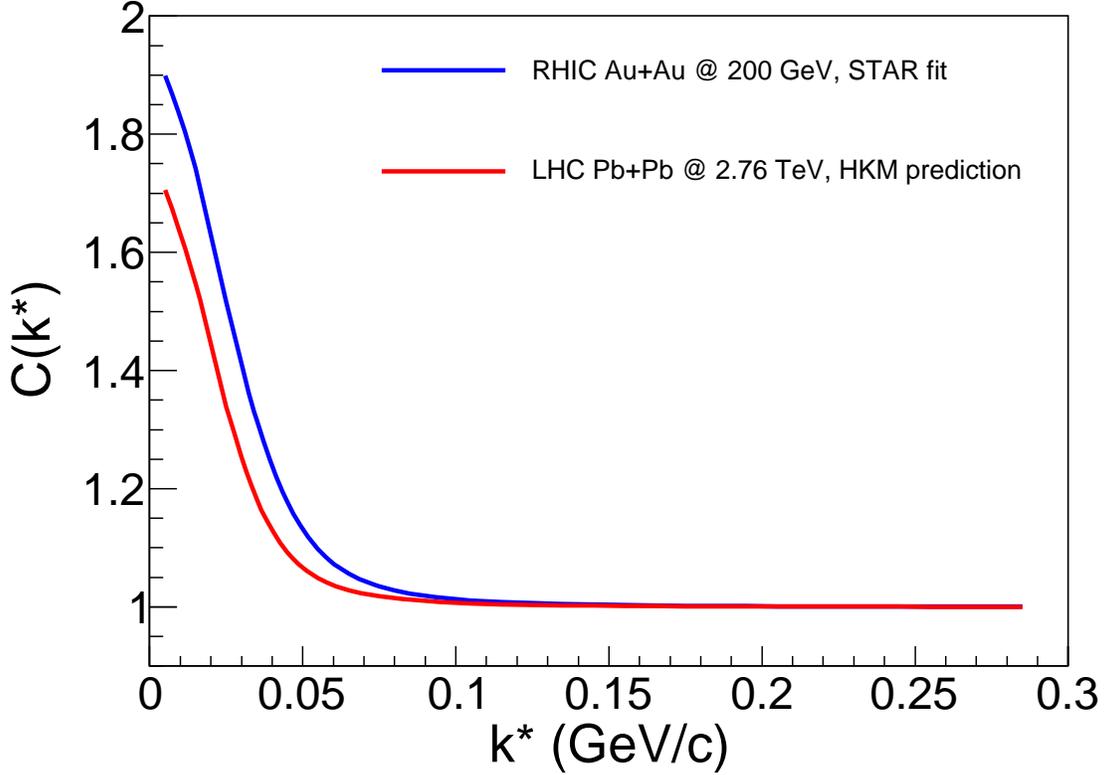}
\caption{The HKM prediction for purity corrected $p-\Lambda \oplus \bar{p}-\bar{\Lambda}$ correlation function in the LHC Pb+Pb collisions at $\sqrt{s_{NN}}=2.76$~TeV, $c=0-5\%$,
$|\eta|<0.8$, with $0.7<p_T<4$~GeV/$c$ for protons and $0.7<p_T<5$~GeV/$c$ for lambdas (red line).
The LHC source radius value calculated in HKM is $r_0=3.76$~fm.
The Lednick\'y-Lyuboshitz fit to the top RHIC energy correlation function, corresponding to
the STAR experiment~\cite{Star}, with $r_0=3.23$~fm extracted from the HKM source function is presented for comparison (blue line).
\label{plamlhc}} 
\end{figure}

Having obtained $r_0$ from HKM and fixing $f^{S}_{0}$ and $d^{S}_{0}$ according to
the paper~\cite{Wang} for baryon-baryon case, we can model the corresponding
correlation function (see Fig.~\ref{plamlhc}). As compared to RHIC, the LHC
$p\Lambda$ CF is slightly more narrow (the two corresponding Gaussian widths differ by a factor of $\sim 1.13$) and has lower intercept of about 1.7.

To build the corresponding baryon-antibaryon correlation function one should
determine the values of parameters entering (\ref{cfunc}) and (\ref{G})
for the LHC case. 
As well as for RHIC we assume $f^{s} = f^{t} = f$ and $d^{s}_{0} = d^{t}_{0} = 0$.
The source radius is again fixed from HKM calculation,
$r_0=3.76$~fm.
As for the real and imaginary parts of the scattering length,
$\Re f_0$ and $\Im f_0$, they characterize antiproton-lambda strong interaction
and hence are not changed when switching from RHIC to LHC. 
So we can use the values extracted from the fit to RHIC $\bar{p}\Lambda$
correlation function shown in Fig.~6 of the paper~\cite{pLamOur},
where the Gaussian parametrization~(\ref{G}) for the residual correlation 
contribution $C_{\mathrm{res}}(k^*)$ is applied, $\Re f_0= 0.14\pm 0.66$~fm and $\Im f_0= 1.53\pm 1.31$~fm.
The parameter $\beta=\tilde{\alpha}\tilde{\beta}$ describes the strength of residual correlations
and the fraction of residually correlated non-primary particles.
So, this parameter depends on particle interaction kinematics. 
Thus, it also can be taken the same as for RHIC, $\beta=0.034\pm 0.005$.
As for the $R$ parameter, being some effective size associated with
residual correlations ($R_{\mathrm{RHIC}}=0.48 \pm 0.05$~fm), one can suppose that for LHC it will be 
larger than for RHIC, approximately proportionally to the source radii 
ratio $r_0^{\mathrm{LHC}}/r_0^{\mathrm{RHIC}}\approx 1.15$, i.~e. $R_{\mathrm{LHC}} = R_{\mathrm{RHIC}}(r_0^{\mathrm{LHC}}/r_0^{\mathrm{RHIC}})$. Such an assumption gives $R_{\mathrm{LHC}}=0.55 \pm 0.06$~fm
\footnote{The calculations show that the model correlation function
depends weakly on $R$ changing by such a close to unity factor.}.
Note, that since the mentioned parameter values have errors, the LHC correlation function
cannot be predicted exactly, but only up to uncertainties caused by these errors
in the fit parameters. The uncertainty in the predicted correlation function is
calculated as 
$\Delta C_{\mathrm{uncorr}}(k^*)=\sqrt{\sum_{i=1}^4 \left(\frac{\partial C_{\mathrm{uncorr}}(k^*)}{\partial x_i} \right)^2\sigma_i^2+2\sum_{1\le i<j\le 4}\frac{\partial C_{\mathrm{uncorr}}(k^*)}{\partial x_i}\frac{\partial C_{\mathrm{uncorr}}(k^*)}{\partial x_j}\sigma_{ij}}$, where $x_i$ are the 4 fit parameters, $\Re f_0$, $\Im f_0$,
$\beta$ and $R$, $\sigma_i = \Delta x_i$ are the corresponding parameter standard
errors, and $\sigma_{ij}$ are the covariances of parameters $x_i$ and $x_j$.

The determination of purity $\lambda(k^*)$ at the LHC is not so explicit,
since it depends not only on the fraction of secondary pairs,
coming from resonance decays, but also on the experimental setup, or more precisely,
on the fraction of misidentified pairs. To clarify this issue we have calculated
the fractions of pairs made by particles having different origination 
(primary or coming from certain decays) in the hydrokinetic model
for RHIC and LHC collisions.
The results obtained in both cases are quite close and are presented in Table~I.
Comparing these fractions with those in Table~III for RHIC~\cite{Star}, one can see
that experimental fraction of $p_{prim}-\Lambda_{prim}$ pairs $\lambda=0.15$ is about
2.5 times lower than in HKM, likely due to misidentification problem, which
takes place in the experiment. In HKM simulations, on the contrary, all the
produced particles are correctly identified, that leads to such a difference
between corresponding fraction values.
However, since HKM is a realistic model that describes both RHIC and LHC 
bulk observables well, basing on its results one can conclude, that true purities,
understood as the fractions of primary pairs, at RHIC and LHC should be quite similar.

\begin{table}
\begin{tabular}{|c|c|}
\hline
Pairs & Fractions (\%) \\ \hline
$p_{prim}-\Lambda_{prim}$                  & $38$ \\ \hline
$p_{\Lambda}-\Lambda_{prim}$                  & $16$ \\ \hline
$p_{\Sigma^{+}}-\Lambda_{prim}$                 & $ 3$ \\ \hline
$p_{prim}-\Lambda_{\Sigma^{0}}$                 & $12$ \\ \hline
$p_{\Lambda}-\Lambda_{\Sigma^{0}}$                 & $ 5$ \\ \hline
$p_{\Sigma^{+}}-\Lambda_{\Sigma^{0}}$                & $ 1$ \\ \hline
$p_{prim}-\Lambda_{\Xi}$                   & $ 17$ \\ \hline
$p_{\Lambda}-\Lambda_{\Xi}$                   & $ 7$ \\ \hline
$p_{\Sigma^{+}}-\Lambda_{\Xi}$                  & $ 1$ \\ \hline
 \end{tabular}
\caption{The fractions of $p\Lambda$ pairs, primary and coming from different decays
calculated in HKM. These fractions are quite similar for RHIC and LHC cases.} 
\end{table}

\begin{figure}
\centering
\includegraphics[bb=0 0 567 409, width=0.9\textwidth]{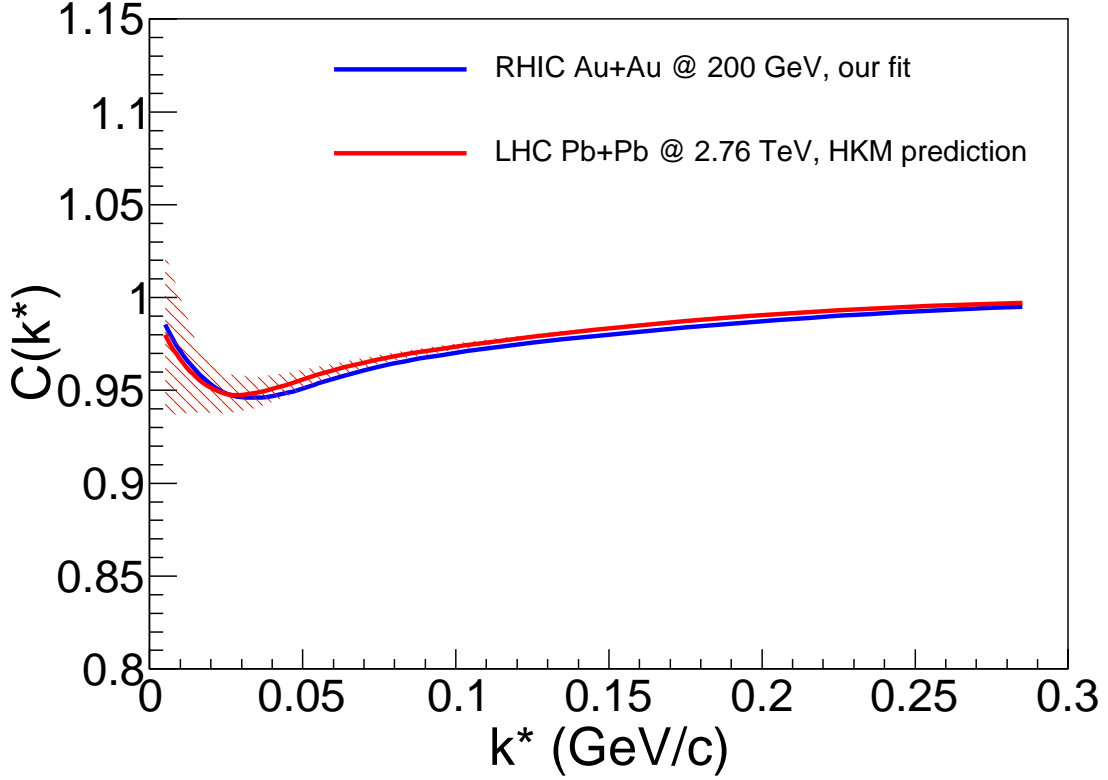}
\caption{The same as in Fig.~\ref{plamlhc} for purity uncorrected $\bar{p}-\Lambda \oplus p-\bar{\Lambda}$ correlation function. 
The HKM source radius for LHC is $r_0=3.76$~fm. The purity $\lambda(k^*)$ is
the same as for RHIC case~\cite{Star}.
The scattering length real and imaginary parts, $\Re f_0$ and $\Im f_0$, are taken from
the fit to RHIC CF that corresponds to Fig.~6 from~\cite{pLamOur},
where HKM source radius $r_0=3.28$~fm and the Gaussian parametrization~(\ref{G}) for the residual correlation contribution $C_{\mathrm{res}}(k^*)$ are utilized.
For the LHC fit the $C_{\mathrm{res}}(k^*)$ parameter $\beta$ coincides with that for RHIC,
while parameter $R$ is scaled by the factor $r_0^{\mathrm{LHC}}/r_0^{\mathrm{RHIC}}$.
The LHC fit is determined up to errors in parameters $\Re f_0$, $\Im f_0$,
$\beta$ and $R$, that is illustrated by the band around the LHC curve. 
 \label{aplamlhc1}}
\end{figure}

\begin{figure}[t]
\centering
\includegraphics[bb=0 0 567 409, width=0.9\textwidth]{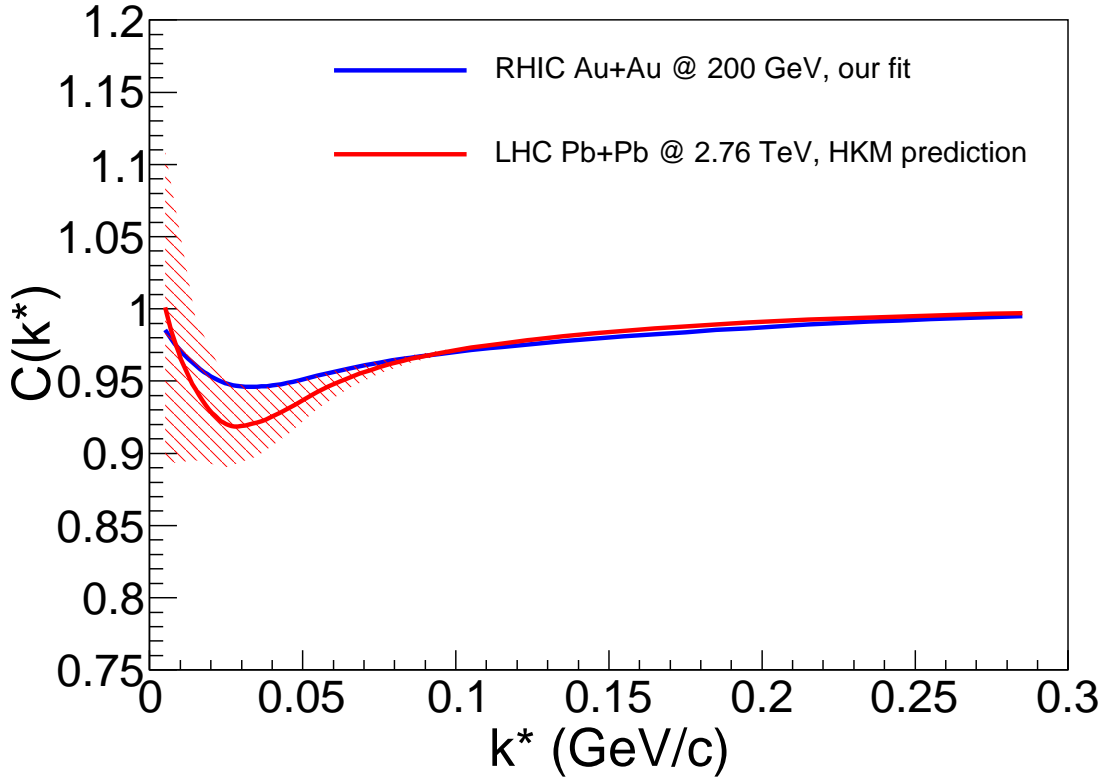}
\caption{The same as in Fig.~\ref{aplamlhc1}, but LHC purity $\lambda(k^*)$
is scaled as compared to RHIC according to primary $p\Lambda$ pairs fraction 
in HKM simulations, where there is no misidentification problem. 
 \label{aplamlhc2}}
\end{figure}

\begin{figure}[t]
\centering
\includegraphics[bb=0 0 567 409, width=0.9\textwidth]{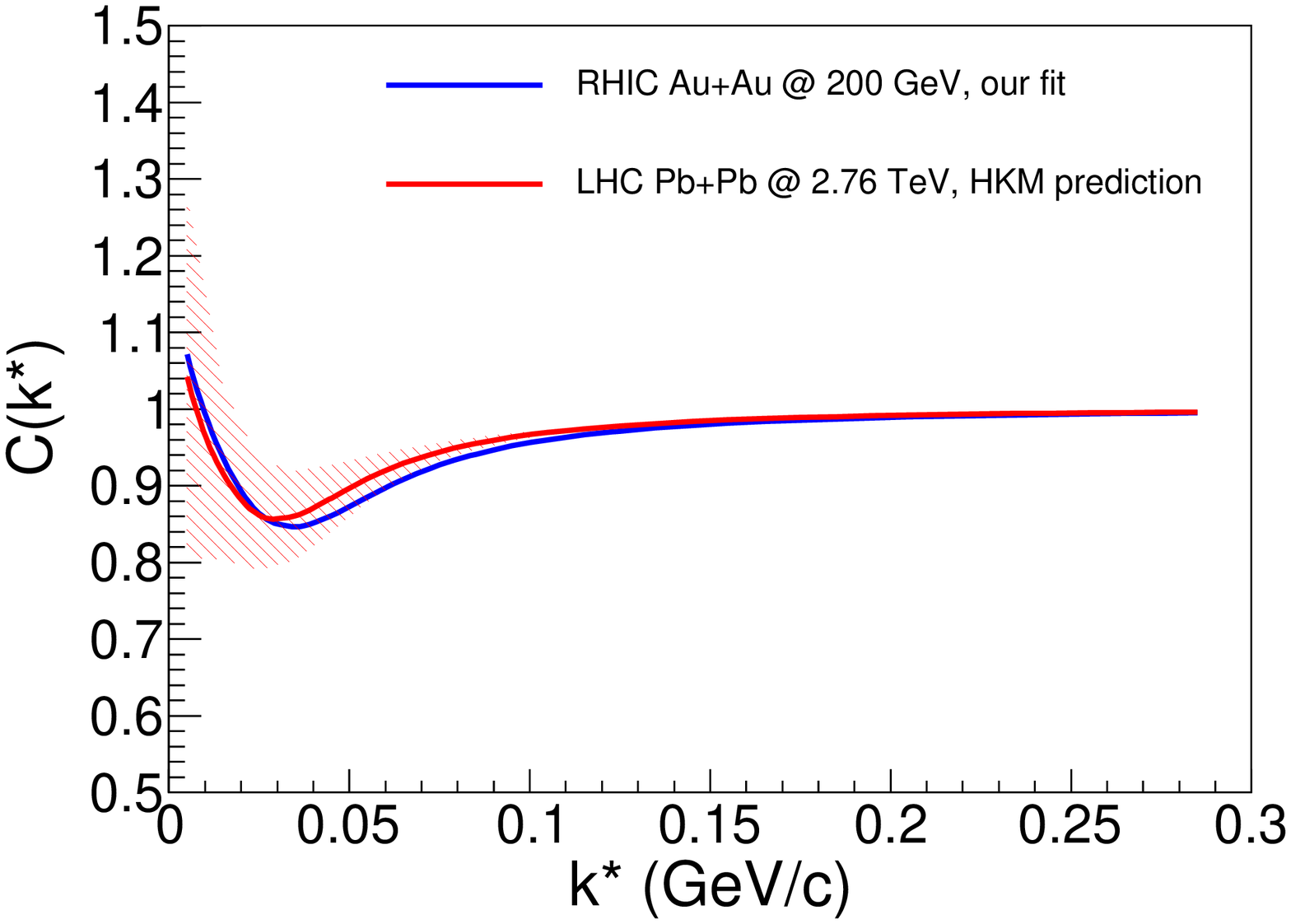}
\caption{The same as in previous two figures, but the correlation
functions are corrected for purity and residual correlations, i.~e. 
$C(k^{*}) = 1 + (C_{\mathrm{uncorr}}(k^{*})-1)/\lambda(k^{*}) - \alpha(k^*) (C_{\mathrm{res}}(k^*)-1)/\lambda(k^{*})$.
 \label{aplamlhc3}}
\end{figure}

In such a situation we demonstrate three different plots for the LHC $\bar{p}-\Lambda \oplus p-\bar{\Lambda}$ correlation function (see Fig.~\ref{aplamlhc1}--\ref{aplamlhc3}).
The first one demonstrates the model CF with $\lambda_{\mathrm{LHC}}(k^*)$ the same as for RHIC~\cite{Star},
$\lambda_{\mathrm{LHC}}(k^*)=\lambda_{\mathrm{RHIC}}(k^*)$.
This function is again more narrow than for RHIC.
In Fig.~\ref{aplamlhc2} one can see our prediction for the LHC correlation function
with $\lambda_{\mathrm{LHC}}(k^*) = 2.5 \lambda_{\mathrm{RHIC}}(k^*)$, where the factor 2.5 
corresponds to the ratio of corresponding primary pairs' fractions in HKM and in the STAR experiment.
As for Fig.~\ref{aplamlhc3}, it shows the purity and residual correlation corrected CFs for LHC
and RHIC. They are expressed (as it follows from Eq.~(\ref{cfunc})) through the uncorrected ones as $C(k^{*}) = 1 + (C_{\mathrm{uncorr}}(k^{*})-1)/\lambda(k^{*}) - \alpha(k^*) (C_{\mathrm{res}}(k^*)-1)/\lambda(k^{*})$.
The latter ``true'' function can be easily compared with the experimental result (corrected in the same way for purity and residual correlations),
since it does not depend on the fraction of misidentified particles in the concrete experiment.
As compared to RHIC, the LHC curve apparently has smaller amplitude and width.

\section{Conclusions}

The first predictions for $p\Lambda$ and $\bar{p}\Lambda$ correlation functions, 
calculated within Lednick\'y-Lyuboshitz and hydrokinetic (HKM) models and accounting for the residual 
correlation effect, are presented for the 5\% most central Pb+Pb
LHC collisions at the energy $\sqrt{s_{NN}}=2.76$~TeV.
The functions' behavior is predicted based on the results previously obtained for 
Au+Au collisions at the top RHIC energy.

Both $p\Lambda$ and $\bar{p}\Lambda$ correlation function curves in the LHC case are slightly more narrow than those at RHIC.

The LHC source radii $r_0$, calculated in HKM for baryon-baryon and baryon-antibaryon cases
are similar, $r_0=3.76$~fm. They are about 1.15 times larger than the corresponding RHIC radii.

The pair purities (the fractions of pairs consisting of primary particles), calculated in HKM for 
RHIC and LHC are very close and are about 2.5 times larger than the experimental primary pairs fraction
in the STAR Collaboration experiment at the RHIC. This difference is most probably because
of particle misidentification problem existing in the experiment.
That is why we present results for purity uncorrected $\bar{p}\Lambda$
correlation functions in two variants: with purity $\lambda(k^*)$ similar to
that in the experiment at RHIC and with RHIC purity scaled by a factor of 2.5, that
corresponds to HKM purity.
The measured LHC correlation function should lie somewhere between these two limiting possibilities.
Depending on the concrete experimental setup and the related fraction of misidentified
particles, the real curve can be closer either to one or to another variant.

We also demonstrate the purity and residual correlation corrected 
baryon-antibaryon CF which should not depend on the experiment details
and thus can be easily compared with experimental result. This function
for LHC has smaller amplitude than for RHIC.

\begin{acknowledgments}
The authors are grateful to Iurii Karpenko for his assistance with computer code.
The research was carried out within the scope of the EUREA:
European Ultra Relativistic Energies Agreement (European
Research Group: ``Heavy ions at ultrarelativistic energies'')
and is supported by the National Academy of Sciences of
Ukraine (Agreement MVC1-2015).  
\end{acknowledgments}

\end{document}